\documentstyle[12pt]{article}
\begin{document}

\begin{flushright}
ITP-SU-95/05 \\
hep-th/9510120 \\
\vspace{0.3cm}
October, 1995
\end{flushright}

\vspace{2.0cm}

\begin{center}

\bf{QUANTUM MECHANICS VIOLATING EFFECTS TO MASSES OF NAMBU-GOLDSTONE BOSONS
: A LESSON FOR MAJORON} \vspace{1cm}

{\rm Y\={u}ichi Chikashige\footnote{e-mail: chika@ge.seikei.ac.jp} and
Tadashi Kon\footnote{e-mail: kon@ge.seikei.ac.jp}}  \vspace{0.5cm}

{\it Faculty of Engineering, Seikei University, Musashino, Tokyo 180, Japan}

\vspace{2cm}

\begin{abstract}
We study gravitational quantum mechanics violating (QMV) effects to masses
of Nambu-Goldstone bosons, taking majoron as an example. We show a
supersymmetric majoron has either mass of O(keV) for the dimension five
potential or smaller mass for effective  potentials with higher dimensions. We
extend the Dashen's  formula for pseudo Nambu-Goldstone bosons to include
possible effects of QMV.
\end{abstract}
\end{center}

\newpage

\setcounter{page}{2}
\baselineskip=24pt

Majoron is a Nambu-Goldstone boson associated with the spontaneous breaking
down (SSB) of global ${\rm U(1)_{\rm B-L}}$ symmetry \cite{{Majoron},{GR}}.
It was originally introduced to give mass of right-handed neutrino in the
seesaw model \cite{{Majoron},{seesaw}}. Majoron is massless, unless
gravitational interaction is introduced. It has been widely argued that
quantum gravitational interaction should not respect any kind of global
symmetry, since black holes are pointed out to cause information loss
\cite{infloss}. Therefore one can say that quantum mechanics violating
(QMV) effects through creation and successive evapolation of black holes
should give majoron nonvanishing mass. Other Nambu-Goldstone bosons like
pion and axion should also get  through QMV effects such additional masses
which would be relatively small values compared to elecroweak masses
\cite{axionmass}. Majoron is different kind from those pseudo
Nambu-Goldstone particles at this point, for this particle has no anomaly
to generate electroweak mass and the ${\rm U(1)_{\rm B-L}}$ symmetry does
not have the gravitational  anomaly to induce gravitational mass if there
would be gravitational instantons.

This paper is concerned with two ways to describe QMV contribution to
masses of Nambu-Goldstone bosons, dealing with majoron as a special
example. One is the effective potential approach which has been extensively
disscussed so far \cite{{Barr},{ABMS}}. However the other seems to be a bit
novel approach which we call the Dashen's formula with QMV effects. The
latter is based on the interesting proposal given by Ellis, Hagelin,
Nanopoulos and Srednicki \cite{EHNS} to illustrate Hawking's idea on QMV
effects in which a pure initial state of a system evolves into a final
mixed state \cite{Hawking82}.

Rothstein, Babu and Seckel in ref.\cite{Barr} and Akhmedov, Berezhiani,
Mohapatra and Senjanovi\'{c} in ref.\cite{ABMS} wrote down such an
effective potential for majoron in terms of expansion with inverse powers
of the Planck mass, $ M_{\rm pl}$, learning lessons of axion case
\cite{Barr}. Akhmedov et al. examine the dimension five potential which
includes the perturbative term $ V_{\rm pl}(\phi, \sigma)$ with
mass-dimension five in addition to the standard Higgs potential $ V_0(\phi,
\sigma)$  as follows :
\begin{equation}
     V  =  V_0(\phi, \sigma)  + V_{\rm pl}(\phi, \sigma) ,
   \label{eq:eq1}
\end{equation}
\begin{equation}
      V_{\rm pl}(\phi, \sigma)  =  V_1(\sigma)  +  V_2(\phi, \sigma).
\label{eq:eq2}
\end{equation}
Here $\phi$ denotes the standard isodoublet Higgs field whose vacuum
expectation value, $ V\simeq $ 246 GeV, gives the Dirac mass of neutrinos
and $\sigma$ represents the isosinglet one whose vacuum expectation value,
$ V_{\rm BL}$, gives the Majorana mass of the right-handed neutrinos. The
minimum of $ V_0(\phi, \sigma)$ determines $ V_{\rm BL}$ which is the scale
of the violation of B $-$ L number conservation. Then $\sigma$ is written
as
\begin{equation}
      \sigma  =  \frac{V_{\rm BL}  +  \rho}{\sqrt{2}} exp(i
\frac{\chi}{V_{\rm BL}}),                          \label{eq:eq3}
\end{equation}
where $\chi$ is the majoron, and $\rho$ is an isosinglet Higgs scalar.
\begin{equation}
      V_1(\sigma)   =  \alpha_{1}\frac{\sigma^5}{M_{\rm pl}}  +
\alpha_{2}\frac{\sigma^{\ast}\sigma^4}{M_{\rm pl}}  +
\alpha_{3}\frac{{\sigma^{\ast}}^2\sigma^3}{M_{\rm pl}}  +  \rm {h.c.}  ,
\label{eq:eq4}
\end{equation}
and
\begin{equation}
      V_2(\phi,\sigma)   =
\beta_{1}\frac{(\phi^{\dag}\phi)^2\sigma}{M_{\rm pl}}  +
\beta_{2}\frac{(\phi^{\dag}\phi)\sigma^2\sigma^{\ast}}{M_{\rm pl}}  +
\beta_{3}\frac{(\phi^{\dag}\phi)\sigma^3}{M_{\rm pl}}  +  \rm {h.c.} .
\label{eq:eq5}
\end{equation}
are the dimension five forms written in ref. \cite{ABMS}.
According to the relative magnitude between $ V$ and $ V_{\rm BL}$, ref.
\cite{ABMS} classifies the two cases (A) and (B).
$ V < V_{\rm BL}$ corresponds to the case (A) where the mass of majoron, $
m_\chi$, becomes approximately
\begin{equation}
      m_\chi  \simeq \sqrt{\beta_{1}}(\frac{V}{V_{\rm BL}})^\frac{1}{2}
\rm{keV}, \label{eq:eq6}
\end{equation}
while $ V_{\rm BL} < V $ is the case (B) and here
\begin{equation}
      m_\chi  \simeq \sqrt{\frac{29}{3}\alpha_{1} + \frac{9}{2}\alpha_{2} +
\frac{1}{2}\alpha_{3}}(\frac{V_{\rm BL}}{V})^\frac{3}{2} \rm{keV}.
\label{eq:eq7}
\end{equation}
Akhmedov et al. mentioned the upper bound for $ V_{\rm BL}$ is constrainted
from the cosmological mass density to be 10 \rm{TeV}. But no further strong
arguments were not given to specify the value of $ V_{\rm BL}$ by them.

Now let us turn to see what happens if we look at supersymmetric (SUSY)
version of majoron. Shiraishi, Umemura and Yamamoto argued in detail such a
model \cite{SUY}  and the identical model was independently discussed by
Giudice, Masiero, Pietrini and Riotto around the same time \cite{GMPR}.
Following the notation of ref. \cite{GMPR}, we have the potential of
Minimal Supersymmetric Standard Model (MSSM), $ V_{0}$, after SUSY breaking
as
\begin{equation}
      V_0  =  V_0(H^{0}_{1},H^{0}_{2})  +  V_0(N,\Phi)  +
V_0(\nu,N,\Phi,H^{0}_{1},H^{0}_{2}).    \label{eq:eq8}
\end{equation}
In this equation, $ V_0(H^{0}_{1},H^{0}_{2})$ is the usual MSSM Higgs
potential, and $ V_0(N,\Phi)$ and $ V_0(\nu,N,\Phi,H^{0}_{1},H^{0}_{2})$
are the terms which are both responsible to break ${\rm U(1)_{\rm B-L}}$
symmetry and $ R$-parity as well. The soft SUSY breaking masses included in
$V_{0}$ are supposed to be an order of 1 TeV as usual. Then as noted by
both groups of the authors of refs. \cite{SUY} and \cite{GMPR}, the
consistency requires that $ R$-parity and ${\rm U(1)_{\rm B-L}}$ symmetry
should also be broken down spontaneously at the same order, 1 TeV. Thus we
admit $ V_{\rm BL} \sim O(1 \rm{TeV})$ for the SUSY majoron. Now the case
(B) in ref. \cite{ABMS} should be chosen for our SUSY majoron and its mass
is said by the above mentioned effective potential approach to QMV effects
of ref. \cite{ABMS} to be an order of keV.  Actually Berezinsky and Valle
expected that a very weakly interacting keV majoron is considered to be a
good candidate for a dark matter particle \cite{BV}.

Then a fundamental question is arisen why the dimension five effective
potential could be more important than other effective potentials with
higher dimensions for our majoron. Let us take an effective potential $
V_{n}$ with an arbitrary dimension $n$ for the case (B) as follows:
\begin{equation}
      V_{n}  =  \alpha_{n}\frac{\sigma^n}{M_{\rm pl}^{n-4}}
\label{eq:eq9}
\end{equation}
This  gives majoron such mass as
\begin{equation}
       m_\chi  =  \frac{(n +4)(n + 3)}{2^{\frac{n}{2}}}
\alpha_{n}(\frac{V_{\rm BL}}{M_{\rm pl}})^{\frac{n}{2}}V_{\rm BL}.
\label{eq:eq10}
\end{equation}
Our SUSY majoron would obtain such an order of mass for each dimension $ n$.
\begin{equation}
       m_\chi \simeq \left\{ \begin{array}{ll}
             10^{-5} \rm {GeV}      & \mbox{for dim 5 ($n$ = 1)} \\
             10^{-13} \rm {GeV}    & \mbox{for dim 6 ($n$ = 2)} \\
             10^{-21} \rm {GeV}    & \mbox{for dim 7 ($n$ = 3)} \\
                             \vdots                  & .
                               \end{array}
                           \right.                  \label{eq:eq11}
\end{equation}
At this stage we don't have any motivations forcing us to choose a special
value of mass among the above. What one can say at most is only that the
origin of mass of majoron should be the QMV effects.

Now let us turn to an alternative approach. Hawking pointed out the fact
that creation and evapolation of black holes let a system loose quantum
coherence \cite{Hawking75}. He then tried to present axioms suitable to
quantum theory of gravity and construct the superscattering operator to
represent loss of quantum coherence \cite{Hawking82}. Following his idea,
Ellis et al. proposed a special form of a differential equation for a
density matrix $\rho$ which describes evolution of a system from a pure
state to a mixed state \cite{EHNS}. Although Banks, Susskind and Peskin
wrote a paper in which this differential equation might cause either
breakdown of causality or violation of energy-momentum
conservation\cite{BSP}, Unruh and Wald have recently published a paper in
which they argue such undesirable features would hardly been seen in our
laboratories \cite{UW}. We are going to follow this viewpoint of Unruh and
Wald.

The equation for $\rho$ written by Ellis et al. \cite{EHNS} is as follows,
according to Unruh and Wald \cite{UW},
\begin{equation}
\dot{\rho} =  - i [H, \rho] -  \sum_{i}^{ } \lambda_{i} (Q_{i} \rho  +
\rho Q_{i} - 2 Q_{i} \rho Q_{i}). \label{eq:eq12}
 \end{equation}
The first term of the right-hand side in eq.(12) is a conventional quantum
mechanical one. The second term of the right-hand side in eq.(12) in which
$ Q_{i}$ is an hermite projection operator,  $ Q_{i}^{\dag} =  Q_{i}$ and
$ Q_{i}^2  = Q_{i}$, implies such a peculiar evolution of the sytem from a
pure state to a mixed state, namely, QMV development. Unruh and Wald have
written the Heisenberg equation with Hamiltonian $ H$ for a Heisenberg
operator $ A_{\rm H}$ in the following form.
\begin{equation}
\dot{A_{\rm H}}  =    i [H, A_{\rm H}] - \sum_{i}^{ } \lambda_{i} (Q_{i}
A_{\rm H}  +  A_{\rm H} Q_{i}  -  2 Q_{i} A_{\rm H} Q_{i})
\end{equation}
\begin{equation}
\mbox{} =  i [H, A_{\rm H}] +\sum_{i}^{ } \lambda_{i}[Q_{i}, [ A_{\rm H},
Q_{i}]] .
\label{eq:eq13}
\end{equation}
(This was noted by Lindblad \cite{Lind} and Gorini, Frigerio, Verri,
Kossakowski and Sudarshan \cite{GFVKS}.)

Now we recall that mass of pseudo Nambu-Goldstone particle obeys the
Dashen's formula \cite{Dashen};
\begin{equation}
      m^2  = - \frac{1}{f^2} \langle0|[Q_{5},\dot{Q_{5}}]|0\rangle
\label{eq:eq14}
\end{equation}
where $Q_{5}$ is a generator of some global symmetry which would be broken
down spontaneously with decay constant $f$. Therefore we can write such a
formula, using the evolution equation for $ Q_{5}$, as
\begin{equation}
\noindent m^2  =  \frac{i}{f^2} \langle0|[Q_{5},[Q_{5},H]]|0\rangle -
\frac{1}{f^2} \sum_{i}^{ }
\lambda_{i}\langle0|[Q_{5},[Q_{i},[Q_{5},Q_{i}]]]|0\rangle .
\label{eq:eq15}
\end{equation}
The second term in the right-hand side of the above equation represents QMV
contribution to the mass of the Nambu-Goldstone boson. If gravitational
interaction would be neglected, this QMV mass should disappear. Thus one
could expect that either $\{\lambda_{i}\}$ would include suppression
factors of $1/{M_{\rm pl}}^k$ or small values of the matrix elements due to
the presence of the projection operators $\{Q_{i}\}$ which communicate
Hilbert space relating to black holes to Hilbert space in our laboratories
or  both kinds of suppression would be included.
As for majoron, B $-$ L current has no anomaly, so that the first term in
the right-hand side of eq.(16) disappears, contrasted with other pseudo
Nambu-Goldstone particles like pion, axion and so on. Therefore the
generator of B $-$ L symmetry, $ Q_{\rm B-L}$, and the projection operators
{$ Q_{i}$} would give majoron $\chi$ such mass as
 \begin{equation}
      m_{\chi}^2  = - \frac{1}{f^2} \sum_{i}^{ }
      \lambda_{i}\langle0|[Q_{\rm B-L},[Q_{i},[Q_{\rm
B-L},Q_{i}]]]|0\rangle,
      \label{eq:eq16}
\end{equation}
if we follow the argument in ref. \cite{EHNS}. The parameters
{$\lambda_{i}$} should determine an order of magnitude of $ m_{\chi}$.

Ref. \cite{EHNS} mentions an interesting inequality which is said as an
accidental coincidence
\begin{equation}
       \lambda \leq  2\times10^{-21} \rm{GeV}    \label{eq:eq17}
\end{equation}
from long baseline neutron interferometry experiment and
K$^{0}$-$\overline{\rm K^{0}}$ system, where $\lambda$ in their paper plays
essentially the same r\^{o}le as our $\{\lambda_{i}\}$ play. We have again
another accidental coincidence with such a value as $10^{-21}$ GeV in the
previous effective potential with dimension seven in
eq.(\ref{eq:eq11}). Of course we cannot take it too seriously at this
stage. Moreover, there seems to be no reason why we would expect to have a
universal contribution of QMV effects. It should  be noted here, however,
that some physical effects caused by such tiny mass as  $10^{-21}$ GeV may
be feasibly triggered for neutrino oscillations in the case of scalar light
particle \cite{cv}.

Hawking stressed that there shouldn't be any suppression factors with
inverse powers of $ M_{\rm {pl}}$ for matrix elements of the scalar
particles in contrast with those of vector bosons and spin 1/2 particles
\cite{Hawking82}. Hawking, Page and Pope once argued furthermore that there
may be even a scalar tachyon \cite{HPP79}. If we would follow this opinion,
we should think doubtfully that the effective potential in
eq.(\ref{eq:eq9}) has  such an suppression factor as ${1/{ M_{\rm
{pl}}^{n-4} } }$. We see an advantage of the approach of the Dashen's
formula on that point, since this formula can be written down in any case
with suppression factors or without them. Certainly one has {\it {a
priori}} no reason to expect non-negative contribution from QMV effects in
the second term of our Dashen's formula, eq.(\ref{eq:eq15}). That means we
need to take definitely much more efforts to examine carefully this vacuum
expectation value of commutators with two generators of a global symmetry
and a couple of projection operators in that term.

In this note we have given a couple of descriptions for masses of
Nambu-Goldstone particles, namely, the effective potential and the Dashen's
formula. For majoron the effective potential approach needs to have the
value of $ V_{\rm BL}$ and a specification of dimension as well in order to
predict the mass. SUSY majoron can provide an interesting value of $ V_{\rm
BL}$. The Dashen's formula needs to analyze deeply  the matrix elements of
commutation relations in the second term of the right-hand side in
eq.(\ref{eq:eq15}) and in eq.(\ref{eq:eq16}). Otherwise we would never
understand what kind of physical process would control masses of
Nambu-Goldstone bosons through QMV effects. \\

One of the authors (Y. C.) would like to thank K. Kawarabayashi for his
valuable
comments.

\end{document}